\documentclass[vecphys]{svmult}

\usepackage{makeidx}         % allows index generation
\usepackage{graphicx}        % standard LaTeX graphics tool
\usepackage{multicol}        % used for the two-column index
\usepackage[bottom]{footmisc}% places footnotes at page bottom
\makeindex             % used for the subject index
\begin{document}

\title*{Star Formation and Molecular Gas in AGN}
\author{R. Davies\inst{1}\and
R. Genzel\inst{1}\and
L. Tacconi\inst{1}\and
F. Mueller S\'anchez\inst{1}\and
%S. Friedrich\inst{1}\and
A. Sternberg\inst{2}}

\authorrunning{R. Davies et al.}

\institute{Max Planck Institut f\"ur extraterrestrische Physik, Postfach 1312, 85741, Garching, Germany
\and School of Physics and Astronomy, Tel Aviv University, Tel Aviv 69978, Israel}
\maketitle

\begin{abstract}
We present initial results from a survey of nearby AGN using the near
infrared adaptive optics integral field spectrograph SINFONI. These
data enable us to probe the distribution and kinematics of the gas and
stars at spatial resolutions as small as 0.085$''$, corresponding
in some cases to less than 10\,pc. In this contribution we present
results concerning
(1) the molecular gas in the nucleus of NGC\,1068
and its relation to the obscuring torus; and 
(2) the stars which exist on spatial scales of a
few tens of parsecs around the AGN, the evidence for their remarkably
young age and extreme intensities, and their relation to the AGN.
\end{abstract}

\section{Introduction}
\label{dav:sec:intro}

Between 2002 and 2005 we have observed and analysed the central
regions of 9 AGN with the following aims in mind:
\begin{itemize}
\item
determine the extent, intensity, and history of recent star formation
and its relation to the AGN;
\item
measure the distribution and kinematics of molecular gas, and
understand its relation to the obscuring material;
\item
derive black hole masses from spatially resolved stellar
kinematics -- both to test the $M_{\rm BH}-\sigma_*$ relation for AGN,
and also to compare to reverberation masses, perhaps
yielding constraints on the broad line region geometry.
\end{itemize}

The sample was selected according to several criteria, of which the
primary constraint was that the nucleus should be bright enough for
adaptive optics correction.
In addition, we required that the galaxies be close enough that small spatial
scales can be resolved at the near infrared diffraction limit of an
8-m telescope.
Finally, we wanted the galaxies to be ``well known'' so that
complementary data could be found in the literature.
These criteria were not applied strictly, since some targets were also
of particular interest for other reasons.
Thus, while the sample cannot be considered 'complete', we
nevertheless believe it provides a reasonable cross-section of AGN;
additional AGN may be added once the VLT Laser Guide Star Facility
is commissioned.
The selected targets are listed in Table~\ref{dav:tab:sample}.
All the observations have now been performed and, although work
concerning the molecular gas and coronal line emission is still in
progress, analysis of the nuclear
star forming properties -- the main subject of this contribution -- is
complete. 

\begin{table}[t]
\centering
\caption{AGN sample}
\label{dav:tab:sample}       % Give a unique label
\begin{tabular}{llrllr}
\hline\noalign{\smallskip}
Target & Classification & Dist. & \ \ \ & \multicolumn{2}{l}{Resolution} \\
       &                & (Mpc)    && arcsec & pc \\
\noalign{\smallskip}\hline\noalign{\smallskip}
Mkn 231$^a$  & ULIRG, Sy\,1, QSO   & 170 && 0.176 & 145 \\
IRAS 05189-2524 \ \ & ULIRG, Sy\,1 & 170 && 0.12  & 100 \\
NGC 2992 & Sy\,1                   &  33 && 0.30  &  48 \\
NGC 3783 & Sy\,1                   &  42 && 0.18  &  37 \\
NGC 7469$^b$ & Sy\,1               &  66 && 0.085 &  27 \\
NGC 1097 & LINER, Sy\,1            &  18 && 0.245 &  21 \\
NGC 3227$^c$ & Sy\,1               &  17 && 0.085 &   7 \\
NGC 1068 & Sy\,2                   &  14 && 0.085 &   6 \\
Circinus$^d$ & Sy\,2               &   4 && 0.22  &   4 \\
%NGC 1316$^*$ & LINER & 25 && Oct '05 && VLT / SINFONI \\ 
\noalign{\smallskip}\hline
\end{tabular}

Detailed studies of individual objects already published: 
$^a$\,Davies et al. 2004a \cite{dav:dav04a};
$^b$\,Davies et al. 2004b \cite{dav:dav04b};
$^c$\,Davies et al. 2006 \cite{dav:dav06}.
$^d$\,M\"uller Sanchez et al. 2006 \cite{dav:mul06}.
\end{table}

\section{Molecular Gas in NGC\,1068}
\label{dav:sec:n1068gas}

\begin{figure}[t]
\centering
\includegraphics[height=5cm]{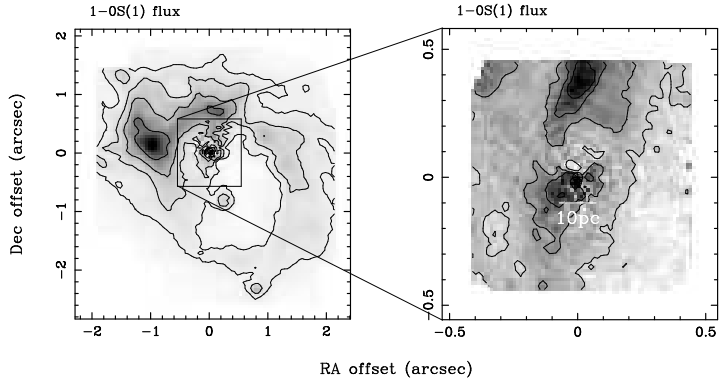}
\caption{Map of H$_2$ 1-0\,S(1) line emission in NGC\,1068 at a
  resolution of 0.085$''$. The AGN is at (0,0) offset.
There is clear evidence for H$_2$ at this
  position with a size scale of 10--20\,pc, which we
  identify with the obscuring molecular torus.}
\label{dav:fig:n1068h2map}       % Give a unique label
\end{figure}

\begin{figure}[t]
\centering
\includegraphics[height=5cm]{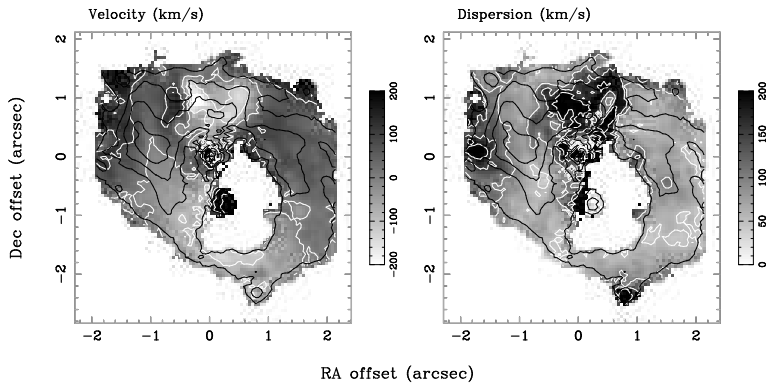}
\caption{The kinematics of the H$_2$ 1-0\,S(1) emission in NGC\,1068
  are far more complex than those of the stars, which have
  PA=85$^\circ$ throughout the field here.
}
\label{dav:fig:n1068h2kin}       % Give a unique label
\end{figure}

As the  cornerstone of AGN unification models, NGC\,1068 is one of the 
most studied objects in the night sky, yet many aspects of its nuclear 
region remain poorly understood, despite its proximity.
In particular, the peculiar gas kinematics revealed in
0.7$''$ CO(2--1) mapping were modelled as a warped disk
\cite{dav:sch00}.
It is now apparent from our 2.12\,$\mu$m H$_2$ 1-0\,S(1) maps of the
distribution 
(Fig.~\ref{dav:fig:n1068h2map}) and kinematics (Fig.~\ref{dav:fig:n1068h2kin}),
which have a spatial resolution nearly 10 times better, that this
picture is too simple (Mueller S\'anchez et al. in prep).

What was previously thought to be a ring of gas at an inner Lindblad
resonance we now find is comprised of several distinct components.
Besides the bright emission east of the AGN, the ring itself is in
fact centred 0.6$''$ (40\,pc) to the southwest of the AGN. 
Its kinematical major axis coincides with its minor axis --
suggestive of inflow or outflow -- at a position angle of about
$135^\circ$ (significantly different to the $85^\circ$ of the stars).
The offset centre and the fact that 1.64\,$\mu$m [Fe{\sc ii}] emission
is seen along the inside edge of the H$_2$ ring suggests that it is
outflowing.
Assuming it is intrinsically circular, the geometry implies an outflow
speed of 150\,km\,s$^{-1}$. 
In addition, the diffuse knot of H$_2$ near the middle of the ring
is moving at 300\,km\,s$^{-1}$ relative to systemic.
It is probable that the momentum and energy needed to drive such
expansion can only be provided by one or more hypernova
\cite{dav:nom05}.

The region north of the AGN is characterised by
double-peaked line profiles (hence the large velocity dispersion),
where the AGN jet has driven its way through the H$_2$.
There is also a linear structure north and south of the AGN.
The tip of the finger extending from the north to within 0.2$''$ of
the AGN coincides with a knot of radio continuum emission
\cite{dav:gal96} -- direct evidence of the shock interface between
the jet and a molecular cloud that has caused the jet direction to
change.
Further on, the jet also brightens where it passes through the
arc of bright 1-0\,S(1) emission to the northeast.

Finally, we have found a knot of H$_2$ emission at the location of
the AGN with a size scale of 10--15\,pc.
Given that this is consistent with the smaller 3\,pc scale of
the 300\,K dust emission \cite{dav:jaf04} as well as the
5--20\,pc scales implied by static torus models
\cite{dav:gra97,dav:sch05,dav:fri06}, we tentatively identify it with
the nuclear obscuring material that is hiding the broad line region.

\section{Nuclear Star Formation in AGN}
\label{dav:sec:starform}

Our AGN sample is rather hetereogeous, including type~1 and~2
Seyferts, ULIRGs, and a QSO.
And their range of distances, coupled with varying adaptive optics
performance, has led to a wide range of spatial resolutions.
Despite this the logarithmic mean resolution of $<$25\,pc is nearly an
order of magnitude smaller than that of other large studies
\cite{dav:cid04}.
And we have, as far as is possible, analysed all the galaxies in a
consistent fashion.
After correcting for dilution by the non-stellar continuum associated
with the AGN, we make use of:
(1) $W_{\rm Br\gamma}$, Br$\gamma$ emission associated with star
formation ratioed to the stellar continuum;
(2) $M/L_K$, the ratio of the mass (estimated where possible using
stellar kinematics) to the K-band stellar luminosity;
(3) $\nu_{\rm SN}$, the supernova rate from spatially resolved data in
the literature, after accounting for a compact AGN contribution.
Davies et al. (in prep) gives details of how these diagnostics are
measured and applied, and assesses the general
properties of nuclear star formation in AGN.
Here we summarise our main results.

%\subsubsection{Nuclear Disks}
\medskip
\noindent
{\bf Nuclear Disks}

\begin{figure}[t]
\centering
\includegraphics[height=4.0cm]{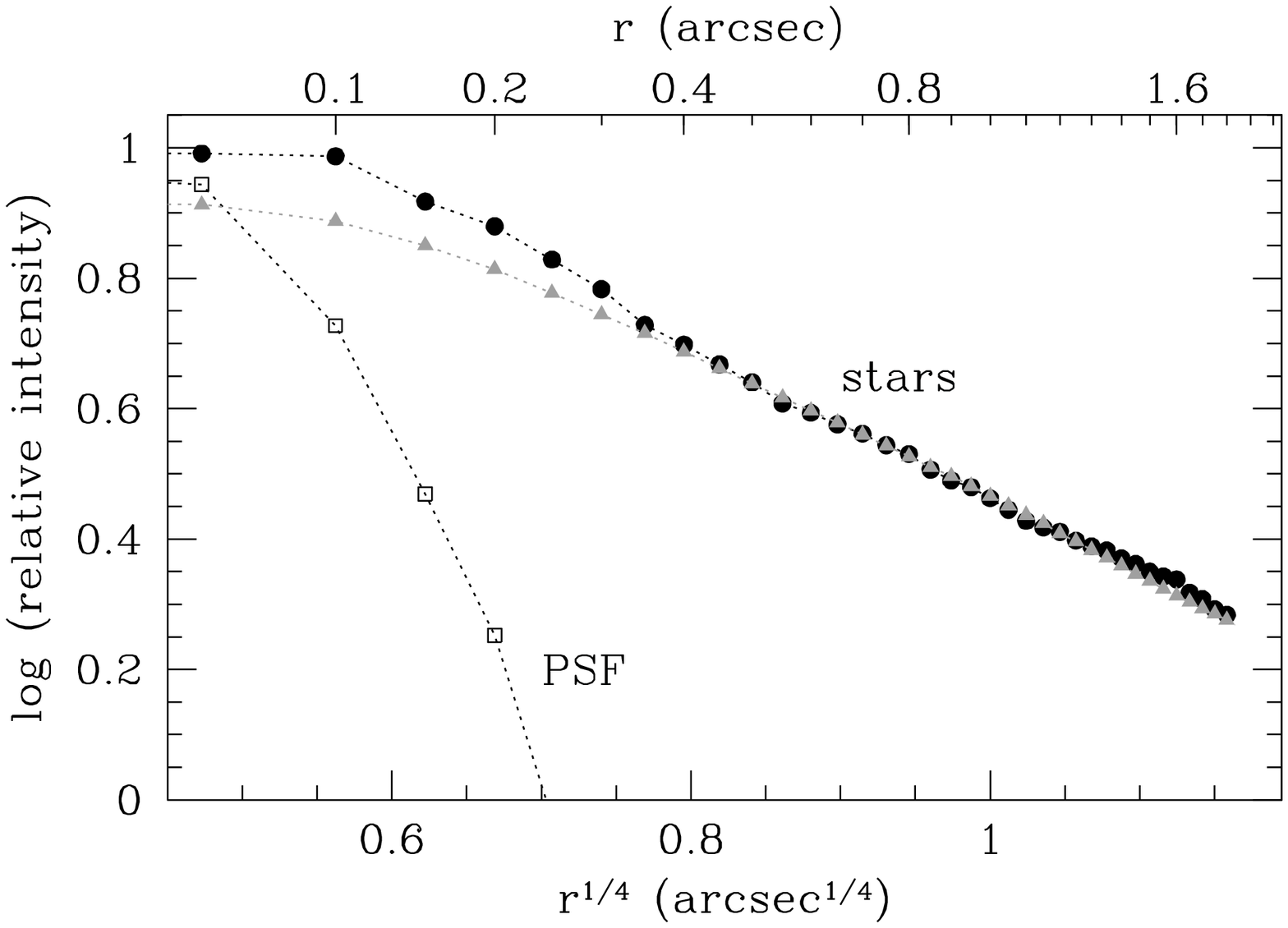}
\includegraphics[height=4.0cm]{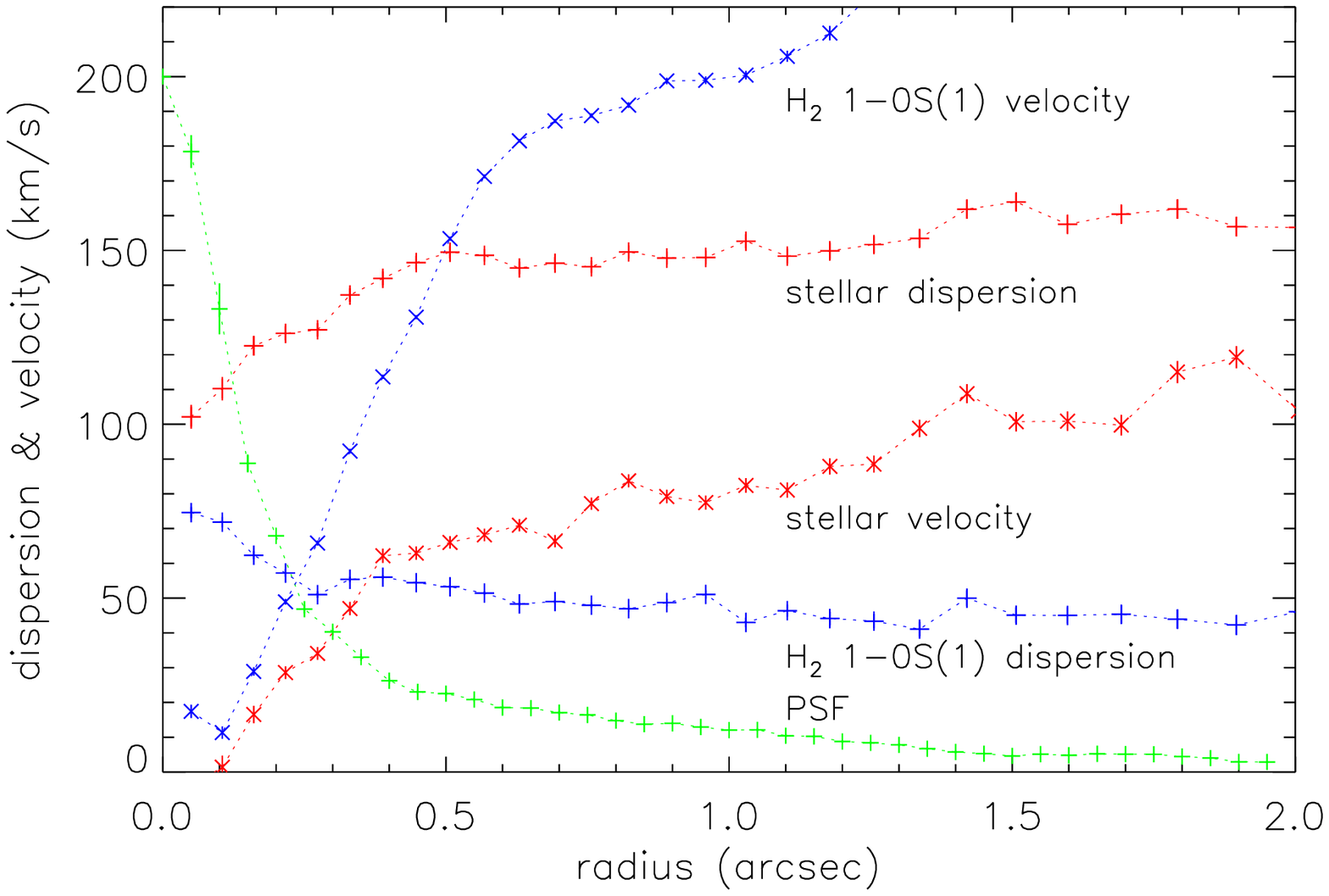}
\caption{Left: azimuthally averaged stellar continuum profile of
  NGC\,1097 (circles), with an $r^{1/4}$ fit extrapolated to the
  centre (triangles).
Right: 1D kinematics of the stars and H$_2$ 1-0\,S(1) in NGC\,1097,
  extracted from the 2D field using kinemetry. 
The radial profile of the PSF (arbitrary flux scaling) is shown for
  comparison.}
\label{dav:fig:n1097}       % Give a unique label
\end{figure}

Tracing the stellar absorption features (e.g. CO bandheads in H- and
K-bands) rather than the broad band continuum, we
have resolved a stellar population close around the AGN in every case:
the stellar intensity increases towards the nucleus on scales
of $<$100\,pc (and in a number of cases $<$50\,pc).
In a few objects (e.g. NGC\,1097, Fig.~\ref{dav:fig:n1097}), we have been
able to show explicitly that there is 
excess stellar continuum in the nucleus above that predicted by an
extrapolation of an $r^{1/4}$ law from the surrounding region.

The age of the star formation in these regions comes primarily from
$\nu_{\rm SN}$ (typically high) and $M/L_K$ (typically low), and lies
in the range 10--300\,Myr.
In most cases we can only estimate a `characteristic' age, since there
may simultaneously be two or more stellar populations that are not
co-eval.
For example, contamination by the bulge population, although having
little effect 
on $W_{\rm Br\gamma}$, could strongly impact $M/L_K$, driving up the
inferred age.
Intriguingly, although the star formation is relatively recent, the
typically low $W_{\rm Br\gamma}$ indicates that there is none actually
on-going, implying that the star formation occurs in relatively short
episodes.

In two cases there is also direct kinematical evidence for young
stars through a spatially resolved  reduction in the stellar
velocity dispersion (on the same
$\sim$50\,pc scale as the excess continuum).
This has been interpreted as arising from stars which have formed
recently from a dynamically cold gas disk, and hence which are
themselves dynamically cool \cite{dav:ems01}.
In NGC\,1097 (Fig.~\ref{dav:fig:n1097}) this coincides with a
change in the kinematics of the molecular gas, which, at radii
$<0.5$$''$, become
increasingly similar to the stellar kinematics.

In conclusion, we have found evidence for recent but short-lived star
formation occuring on scales of $\sim$50\,pc that is dynamically
cooler than the bulge: strong support for the existence of nuclear
disks.

%\subsubsection{Eddington Limited Starbursts}
\medskip
\noindent
{\bf Eddington Limited Starbursts}

When active, the star formation rate on scales of 0.1--1\,kpc is of
order 10--50\,M$_\odot$\,yr$^{-1}$\,kpc$^{-2}$, increasing to 
$\sim$300\,M$_\odot$\,yr$^{-1}$\,kpc$^{-2}$ on scales of 10\,pc. 
Consequently the stellar luminosity close around the AGN is, per unit
area, very high.
In contrast the area itself is small, so that within a few tens of
parsecs the luminosity due to star 
formation is only a few percent of that of the AGN;
but on scales of 1\,kpc, the two can often be comparable.

\begin{figure}[t]
\centering
\includegraphics[height=5.5cm]{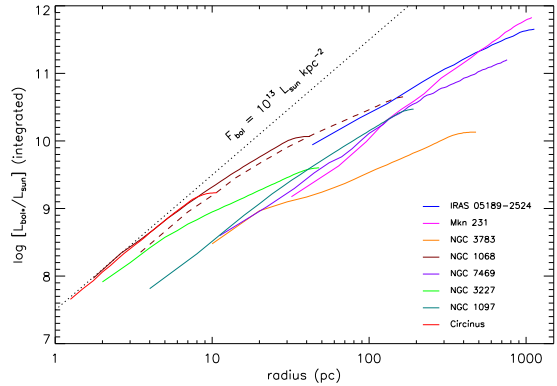}
\caption{Stellar bolometric luminosity per unit area as a function of
  distance from the AGN, showing that in general the stellar intensity
  increases towards $10^{13}$\,L$_\odot$\,kpc$^{-2}$ in the central
  few parsec.}
\label{dav:fig:magbol}       % Give a unique label
\end{figure}

Fig.~\ref{dav:fig:magbol} shows the integrated stellar bolometric luminosity
$L_{\rm *bol}$ as a function of radius.
This figure is robust, despite assuming that all the K-band
light is associated with young stars, because the vertical
axis covers nearly 5 orders of magnitude while the ratio 
$L_{\rm *bol}/L_K$ varies by no more than $\pm$0.3\,dex for
almost any star formation scenario.
It is remarkable that the curves all roughly follow the same trend,
approaching $10^{13}$\,L$_\odot$\,kpc$^{-2}$ on the smallest scales.
Both the luminosity density limit and the dependence on radius are
similar to those for optically thick starburst disk models of ULIRGs
\cite{dav:tho05}.
In these models, ULIRGs radiate at the the
Eddington limit for a starburst, defined as when the radiation
pressure dominates over self-gravity.
Radiation pressure should halt further accretion for
star clusters with luminosity-to-mass ratios exceeding
$\sim500$\,L$_\odot$/M$_\odot$, which occurs when the upper end of the
mass function becomes populated \cite{dav:sco03}.
Crucially, even if stars form as fast as gas is fed in, this limit
can only be exceeded for $\sim$30\,Myr, and less if the efficiency is
lower or some of the gas was present at the start.
This may explain why we see no evidence for current active star
formation: starbursts can only be sustained at this intensity for
short periods.
We suggest that starbursts in AGN are episodic phenomena, proceeding
in a sequence of short and very intense bursts.
Whether one finds evidence for recent star formation in an AGN would
therefore depend on how much time has passed since the most recent
episode.

%\subsubsection{Starburst-AGN Connection}
\medskip
\noindent
{\bf Starburst-AGN Connection}

Given that the star formation we have discussed here occurs on scales
of $<$50\,pc, it is inevitable that it and the AGN will have some
mutual influence on each other.
While the exact form remains elusive, our data do hint at a
possible relationship between the characteristic age of the star
formation and the accretion rate onto the AGN.
The AGN which are radiating
at lower efficiency $<0.1$\,L/L$_{\rm Edd}$ are associated with
starbursts younger than 50--100\,Myr;
AGN that are accreting and radiating more efficiently
$>0.1$\,L/L$_{\rm Edd}$ have starbursts older than 50--100\,Myr.
This implies that there could be a delay between starburst activity and
AGN activity.
What might cause such a delay is beyond the scope of this
contribution, and will be discussed in a future paper.

% BibTeX users please use
% \bibliographystyle{}
% \bibliography{}
%
% Non-BibTeX users please follow the syntax
% the syntax of "referenc.tex" for your own citations

%%%%%%%%%%%%%%%%%%%%%%%%%%%%%%%%%%%%%%%%%%%%%%%%%%%%%%%%%%%%%%%%%%%%%%  }

%%%%%%%%%%%%%%%%%%%%%%%%%%%%%%%%%%%%%%%%%%%%%%%%%%%%%%%%%%%%%%%%%%%%%%

\printindex
\end{document}